\documentclass[12pt]{article}

\author{Yu.~M.~Zinoviev
       \thanks{E-mail address: Yurii.Zinoviev@ihep.ru} \\
        {\it Institute for High Energy Physics} \\
        {\it of National Research Center "Kurchatov Institute"} \\
        {\it Protvino, Moscow Region, 142280, Russia}}
\title{Massive two-column bosonic fields \\
in the frame-like formalism}

\date{}

\begin{document}

\setlength{\unitlength}{1mm}

\maketitle

\begin{abstract}
In this paper we develop the frame-like gauge invariant formulation for
the massive two-column bosonic fields in (anti) de Sitter space-times.
We begin with the partially massless cases in $AdS$ and $dS$ and then
we combine these results into the general massive theory. Separate
section is devoted to the special case where both columns have equal
number of indices.
\end{abstract}

\thispagestyle{empty}
\newpage
\setcounter{page}{1}

\section*{Introduction}

Two-column fields (tensors with two groups of completely anti-symmetric
indices) are the special and interesting case of the general
mixed symmetry fields that naturally arise in space-time dimensions
greater than four. One of the reason for this interest is that such
fields (in the appropriate dimensions) can be considered as the dual
forms of the usual spin-2 graviton (see e.g. \cite{BP12,BBB15} and
references therein). In the frame-like formalism the general massless
mixed symmetry fields (including two-column ones) in the flat
Minkowski space have been constructed in \cite{Skv08}. One of the
specific features of the massless mixed symmetry fields is that most
of them do not admit straightforward deformation into (anti) de
Sitter space and to keep all the gauge symmetries one has to introduce
some additional fields \cite{BMV00} (see also \cite{BIS08,BIS08a}).
Thus the irreducible $(A)dS$ representations for such fields become
reducible ones in the flat limit. In what follows by massless fields
we will always assume massless fields in the Minkowski space while all
other cases that become reducible in the flat limit we will call
partially massless or massive. Partially massless two-column fields in
$AdS$ space have been constructed in \cite{Alk03} as the special case
of the general formalism in \cite{ASV03,ASV05,ASV06}. Note that
approaches of \cite{Skv08} and \cite{Alk03} are quite different. In
the first one to describe two-column fields $Y(k,l)$, $k > l$, one
introduces physical $k$-form with $l$ anti-symmetric local indices
$e^{a[l]}$ and auxiliary $l$-form with $k$ local ones $\omega^{a[k]}$.
At the same time in the second approach both physical and auxiliary
fields are $l$-forms with $k$ and $k+1$ local indices correspondingly.
Let us stress that the second approach keeps only half of the gauge
symmetries that massless two-column fields possess in the flat case
and in what follows we will see that the results of \cite{Alk03} can
be reproduced by the partial gauge fixing.

In the metric-like formalism the general massive two-column fields in
the $(A)dS$ spaces with all possible (partially) massless limits have
been constructed in \cite{Med03} using so-called multi-forms (see e.g.
\cite{MH03} and references therein).\footnote{BRST approach for the
general massive mixed symmetry fields have been developed in
\cite{BKT07,BR11} (see also \cite{Res14,Res16} for the two-column
case), while frame-like gauge invariant description for the massive
two-raw fields were given in \cite{Zin08c,Zin09c}. Weyl actions and
their possible $AdS$ and massive deformations for the two-column
fields have been considered recently in \cite{JM16}.} Our aim here is
to reconstruct the results of \cite{Med03} in the frame-like gauge
invariant formalism following the approach of \cite{Skv08}. As it
could be expected the results of \cite{Med03} appeared to be quite
similar to the results for the simplest "hook" and "window" cases
\cite{Zin02a}. So to simplify calculations we begin with the partially
massless cases in de Sitter and anti de Sitter spaces and than we
combine them into the general massive one.

Our main objection here is that the frame-like formalism seemed to be
not only much simpler but also very effective in the investigations of
possible interactions. Till now the results on the mixed symmetry
fields interactions are not numerous. There was a number of "no-go"
results (e.g. \cite{BBH02,BC04,BBc04}). In \cite{Alk10} cubic
interactions for the system containing not only completely symmetric
fields but also the "long hook" ones in $AdS_5$ have been constructed.
Gravitational interactions for the simple hook both in Minkowski and
in $AdS$ space were considered in \cite{BSZ11}, while a complete
non-trivial theory for the set of "high hooks" have been constructed
in \cite{BS11}. For the massive hook electromagnetic and gravitational
interactions have been considered in \cite{Zin10a,Zin11}.

The paper is organized as follows. In the first section we provide all
necessary formulas for the frame-like description of massless
two-column fields in Minkowski space following \cite{Skv08}. Then in
sections 2 and 3 we consider partially massless cases in anti de
Sitter and de Sitter spaces combining them in section 4 into general
massive theory. Section 5 devoted to the special case $Y(k,k)$ where
both columns have equal number of cells.

\noindent{\bf Notation and conventions} We will work in the (anti) de
Sitter space with the background frame $e^a$  and its inverse 
$\hat{e}_a$. We will heavily use short-hand notation for their wedge
products:
$$
E^{a(k)} = e^{a_1} \wedge e^{a_2} \wedge \dots \wedge e^{a_k}
$$
$$
\hat{E}_{a(k)} = \hat{e}_{a_1} \wedge \hat{e}_{a_2} \wedge \dots
\wedge \hat{e}_{a_k}
$$
A couple of useful relations:
$$
\hat{E}_{a(k)} \wedge e^b = \delta_a{}^b \hat{E}_{a(k-1)} 
$$
$$
\hat{E}_{a(k)} \wedge e^a = (d-k+1) \hat{E}_{a(k-1)}
$$
An $(A)dS$ covariant derivative $D$ is defined so that
$$
D \wedge D \xi^a = - \kappa E^a{}_b \xi^b
$$
In what follows we will systematically omit the wedge product sign
$\wedge$.

\section{Massless fields in the Minkowski space}

As we have already noted, massless fields in the Minkowski space play
the role of minimal building blocks for the construction of general
partially massless and massive fields in $(A)dS$ spaces with arbitrary
value of the cosmological constant. So in this section we give all
necessary formulas for the massless two-column fields. We will follow
the approach of Skvortsov \cite{Skv08} but we will use an
anti-symmetric basis which is natural for the two-column fields.

\subsection{General case $Y(k,l)$, $k > l$}

Frame-like description requires physical $k$-form $R_k^{a(l)}$ as well
as auxiliary $l$-form $\Sigma_l^{a(k+1)}$. The free Lagrangian can be
written as follows:
\begin{equation}
{\cal L}_0(\Sigma_l^{k+1}, R_k^l) = a_0(k,l) \hat{E}_{a(2l)}
\Sigma_l^{a(l)b(k-l+1)} \Sigma_l^{a(l)}{}_{b(k-l+1)} + 
\hat{E}_{a(k+l+1)} \Sigma_l^{a(k+1)} d R_k^{a(l)} \label{massless}
\end{equation}
$$
a_0(k,l) = (-1)^{k-l} \frac{(k+1)!}{2l!}
$$
where $d$ is a usual external derivative. This Lagrangian is invariant
under the following local gauge transformations:
\begin{equation}
\delta_0 \Sigma_l^{a(k+1)} = d \eta_{l-1}^{a(k+1)}, \qquad
\delta_0 R_k^{a(l)} = d \xi_{k-1}^{a(l)} + E_{b(k-l+1)} 
\eta_{l-1}^{a(l)b(k-l+1)}
\end{equation}
where $\xi_{k-1}^{a(l)}$ is a $(k-1)$-form, while
$\eta_{l-1}^{a(k+1)}$ is a $(l-1)$-form.

Note that such massless field does not admit deformation into $(A)dS$
space without introduction of some additional fields (see below). In
this particular case (see general discussion in \cite{BMV00}) the
reason is simple: we can not add to the Lagrangian any term quadratic
in the physical field $R_k^{a(l)}$.

\subsection{Special case $Y(k,k)$}

In this case both physical field $R_k^{a(k)}$ as well as auxiliary one
$\Sigma_k^{a(k+1)}$ are $k$-forms. The free Lagrangian looks like (it
corresponds to (\ref{massless}) for $l=k$):
\begin{equation}
{\cal L}_0 (\Sigma_k^{k+1}, R_k^k) = \frac{(k+1)}{2} \hat{E}_{a(2k)}
\Sigma_k^{a(k)b} \Sigma_k^{a(k)}{}_b + \hat{E}_{a(2k+1)} 
\Sigma_k^{a(k+1)} d R_k^{a(k)}
\end{equation}
and is invariant under the following gauge transformations:
\begin{equation}
\delta_0 \Sigma_k^{a(k+1)} = d \eta_{k-1}^{a(k+1)}, \qquad 
\delta_0 R_k^{a(k)} = d \xi_{k-1}^{a(k)} + e_b \eta_{k-1}^{a(k)b}
\end{equation}
where both gauge parameters are $(k-1)$-forms.

One of the reasons why this case is indeed special is that this
massless field admits deformation into $AdS$ space without
introduction of any additional fields. Let us consider the following
deformed Lagrangian:
\begin{equation}
{\cal L}_0 = \frac{(k+1)}{2} \hat{E}_{a(2k)} \Sigma_k^{a(k)b}
\Sigma_k^{a(k)}{}_b + \hat{E}_{a(2k+1)} \Sigma_k^{a(k+1)} D
R_k^{a(k)} + b_1 \hat{E}_{2k} R_k^{a(k)} R_k^{a(k)}
\end{equation}
where $D$ is $AdS$ covariant derivative, and corrected gauge
transformations:
\begin{equation}
\delta_0 \Sigma_k^{a(k+1)} = D \eta_{k-1}^{a(k+1)} + \beta_1 e^a
\xi_{k-1}^{a(k)},
\qquad \delta_0 R_k^{a(k)} = D \xi_{k-1}^{a(k)} + e_b 
\eta_{k-1}^{a(k)b}
\end{equation}
Variations under the $\eta$ transformations give
$$
\delta_\eta {\cal L}_0 = k [(k+1)(d-2k)\kappa - 2b_1 ] 
\hat{E}_{a(2k-1)} \eta_{k-1}^{a(k)b} R_k^{a(k-1)}{}_b
$$
while variations under the $\xi$ transformations give
\begin{eqnarray*}
\delta_\xi {\cal L}_0 &=& [ (-1)^k 2b_1 - (k+1)(d-2k)\beta_1 ] 
\hat{E}_{a(2k)} R_k^{a(k)} D \xi_{k-1}^{a(k)} \\
 && + k(k+1)(d-2k) [ (-1)^k \kappa - \beta_1 ] \hat{E}_{a(2k-1)}
\Sigma_k^{a(k)}{}_b \xi_{k-1}^{a(k-1)b}
\end{eqnarray*}
Thus we obtain
\begin{equation}
2b_1 = (k+1)(d-2k)\kappa, \qquad \beta_1 = (-1)^k \kappa
\end{equation}

\section{Partially massless case in anti de Sitter space}

In this section we will show that the combination of two fields
$Y(k,l)$ and $Y(k-1,l)$ can be deformed into anti de Sitter space with
$\kappa < 0$ and such deformation keep all four (appropriately
modified) gauge symmetries. Note that this case have already been
considered in our paper \cite{BSZ11}, so we briefly reproduce
corresponding results in our current notation and conventions because
we will need them for the general massive case.

\subsection{Lagrangian and gauge transformations}

We introduce two pairs of physical and auxiliary fields 
($\Sigma_l^{k+1}$, $R_k^l$) and ($\omega_l^k$, $h_{k-1}^l$). We begin
with the sum of their kinetic terms:
\begin{eqnarray}
{\cal L}_0 &=& {\cal L}_0 (\Sigma_l^{k+1}, R_l^k) + {\cal L}_0
(\omega_l^k, h_{k-1}^l)
\end{eqnarray}
and their initial gauge transformations:
\begin{eqnarray}
\delta_0 \Sigma_l^{a(k+1)} &=& D \eta_{l-1}^{a(k+1)}, \qquad
\delta_0 R_k^{a(l)} = D \xi_{k-1}^{a(l)} + E_{b(k-l+1)}
\eta_{l-1}^{a(l)b(k-l+1)} \nonumber \\
\delta_0 \omega_l^{a(k)} &=& D \eta_{l-1}^{a(k)}, \qquad
\delta_0 h_{k-1}^{a(l)} = D \xi_{k-2}^{a(l)} + E_{b(k-l)}
\eta_{l-1}^{a(l)b(k-l)}
\end{eqnarray}
where all derivatives are $AdS$ covariant ones. Than we add all
possible terms with one derivative:
\begin{equation}
{\cal L}_1 = a_3 \hat{E}_{a(k+l-1)} \Sigma_l^{a(k)}{}_b 
h_{k-1}^{a(l-1)b} + a_4 \hat{E}_{a(k+l)} \omega_l^{a(k)} R_k^{a(l)}
\end{equation}
and corresponding corrections to the gauge transformations
\begin{eqnarray}
\delta_1 \Sigma_l^{a(k+1)} &=& \alpha_5 e^a \eta_{l-1}^{a(k)}, \qquad
\delta_1 R_k^{a(l)} = \alpha_6 e^a e_b \xi_{k-2}^{a(l-1)b} \nonumber
 \\
\delta_1 \omega_l^{a(k)} &=& \alpha_7 e_b \eta_{l-1}^{a(k)b}, \qquad
\delta_1 h_{k-1}^{a(l)} = \alpha_8 \xi_{k-1}^{a(l)}
\end{eqnarray}
To cancel all variations with one derivative $\delta_0 {\cal L}_1 +
\delta_1 {\cal L}_0$ (see Appendix A for details) we have to put:
\begin{equation}
\alpha_5 = - \alpha_6 = - \frac{(-1)^la_4}{(k+1)(d-k-l)}, \quad
\alpha_7 = (-1)^{k-l} a_4, \quad
\alpha_8 = - a_4, \quad
a_3 = (-1)^l l a_4
\end{equation}
Note that in this case case there are no any terms quadratic in the
physical fields $R$ and $h$ that can be added to the Lagrangian. But
all variations without derivatives $\delta_1 {\cal L}_1$ vanish
provided
\begin{equation}
a_4{}^2 = - (k+1)(d-k-l)\kappa
\end{equation}

\subsection{Gauge invariant curvatures}

One of the nice features of the frame-like formalism is that for each
field (physical or auxiliary) one can construct gauge invariant object
(that we will call curvature). It is clear that in any gauge theory
such gauge invariant objects play important role both in the free
theory as well as in any attempt to construct non-trivial
interactions.

For the case at hands we can construct four such gauge invariant
curvatures\footnote{These expressions can be easily read out from the
ones for gauge transformations. Of course, after that one has to check
(as we have done) that the resulting curvatures are indeed gauge
invariant.}:
\begin{eqnarray}
{\cal R}_{l+1}^{a(k+1)} &=& D \Sigma_l^{a(k+1)} + \alpha_5 e^a
\omega_l^{a(k)} \nonumber \\
{\cal T}_{k+1}^{a(l)} &=& D R_k^{a(k+1)} + (-1)^{k-l}
E_{b(k-l+1)} \Sigma_l^{a(l)b(k-l+1)} - \alpha_6 e^a e_b 
h_{k-1}^{a(l-1)b} \nonumber \\
{\cal R}_{l+1}^{a(k)} &=& D \omega_l^{a(k)} + \alpha_7 e_b
\Sigma_l^{a(k)b} \\
{\cal T}_k^{a(l)} &=& D h_{k-1}^{a(l)} - (-1)^{k-l} E_{b(k-l)}
\omega_l^{a(l)b(k-l)} - \alpha_8 R_k^{a(l)} \nonumber
\end{eqnarray}
These curvatures satisfy the following differential identities:
\begin{eqnarray*}
D {\cal R}_{l+1}^{a(k+1)} &=& - \alpha_5 e^a {\cal R}_l^{a(k)} \\
D {\cal T}_{k+1}^{a(l)} &=& - E_{b(k-l+1)} 
{\cal R}_{l+1}^{a(l)b(k-l+1)} - \alpha_6 e^a e_b 
{\cal T}_{k-1}^{a(l-1)b} \\
D {\cal R}_{k+1}^{a(l)} &=& - \alpha_7 e_b {\cal R}_{l+1}^{a(k)b} \\
D {\cal T}_k^{a(l)} &=& - E_{b(k-l)} {\cal R}_{l+1}^{a(l)b(k-l)} -
\alpha_8 {\cal T}_{k+1}^{a(l)}
\end{eqnarray*}
Using these curvatures one can rewrite the free Lagrangian in the
explicitly gauge invariant form:
\begin{eqnarray}
{\cal L} &=& \hat{E}_{a(2l+2)} [ b_1 
{\cal R}_{l+1}^{a(l+1)}{}_{b(k-l)} {\cal R}_{l+1}^{a(l+1)b(k-l)} + b_2
{\cal R}_{l+1}^{a(l+1)}{}_{b(k-l-1)} {\cal R}_{l+1}^{a(l+1)b(k-l-1)} ]
\nonumber \\
 && + b_3 \hat{E}_{a(k+l+1)} {\cal R}_{l+1}^{a(k+1)} {\cal T}_k^{a(l)}
\end{eqnarray}
where
$$
[ \frac{(d-k-l-1)}{(k+1)(d-k-l)} b_1 + b_2 ] = - 
\frac{a_0(k,l)}{(l+1)^2a_4{}^2} , \qquad
b_3 = - \frac{(-1)^l}{a_4}
$$
Here we face an ambiguity in the choice of coefficients $b_{1,2,3}$.
It is related with the following identity:
\begin{eqnarray*}
0 &=& \hat{E}_{a(2l+3)} D [ {\cal R}_{l+1}^{a(l+2)b(k-l-1)} 
{\cal R}_{l+1}^{a(l+1)b(k-l-1)} ] \\
 &=& - (-1)^l (l+2) a_4 \hat{E}_{a(2l+2)} [   
{\cal R}_{l+1}^{a(l+1)b(k-l)} {\cal R}_{l+1}^{a(l+1)b(k-l)} \\
 && \qquad \qquad - \frac{(d-k-l-1)}{(k+1)(d-k-l)} 
{\cal R}_{l+1}^{a(l+1)b(k-l-1)} {\cal R}_{l+1}^{a(l+1)b(k-l-1)} ]
\end{eqnarray*}
where we have used the differential identities given above.

\subsection{Partial gauge fixing}

Relatively large number of components and corresponding curvatures
make investigations of possible interactions for such fields rather
involved. To simplify these investigations one can try to use partial
gauge fixing\footnote{Of course, one can not be sure that these two
procedures, namely switching on an interaction and partial gauge
fixing, commute. For the simplest example of mixed symmetry field
$Y(2,1)$ (the so-called hook) it was shown in \cite{BSZ11} that it is
indeed the case.}. In the case at hands we use $\xi_{k-1}^{a(l)}$
transformations and put $h^{a(l)} = 0$. Then the on-shell constraint
${\cal T}_k^{a(l)} \approx 0$ (which is algebraic in this gauge) can
be solved and gives:
$$
\quad R_k^{a(l)} = \frac{(-1)^{k-l}}{a_4} E_{b(k-l)}
\omega_l^{a(l)b(k-l)}
$$
Taking into account that the field $\omega^{a(k)}$ will play the role
of physical (and not auxiliary) field now, we make a re-scaling:
$$
\omega^{a(k)} \Rightarrow (-1)^{k-l} a_4 \omega^{a(k)}
$$
After that the free Lagrangian takes the form:
\begin{eqnarray}
{\cal L} &=& \frac{(-1)^{k-l}(l+1)}{2} \hat{E}_{a(2l)}
\Sigma_l^{a(l)}{}_{b(k-l+1)} \Sigma_l^{a(l)b(k-l+1)} 
+ \hat{E}_{a(2l+1)} \Sigma_l^{a(l+1)}{}_{b(k-l)} D
\omega_l^{a(l)b(k-l)} \nonumber \\
 && - \frac{(-1)^{k-l}(l+1)(d-k-l)}{2} \kappa \hat{E}_{a(2l)}
\omega_l^{a(l)}{}_{b(k-l)} \omega_l^{a(l)b(k-l)}
\end{eqnarray}
This Lagrangian is still invariant under the remaining gauge
transformations:
\begin{eqnarray}
\delta_0 \Sigma_l^{a(k+1)} &=& D \eta_{l-1}^{a(k+1)} + (-1)^k \kappa
e^a \eta_{l-1}^{a(k)} \nonumber \\
\delta_0 \omega_l^{a(k)} &=& D \eta_{l-1}^{a(k)} + e_b
\eta_{l-1}^{a(k)b}
\end{eqnarray}
Such a procedure leaves us with just two gauge invariant curvatures:
\begin{eqnarray}
{\cal R}_{l+1}^{a(k+1)} &=& D \Sigma_l^{a(k+1)} + (-1)^k \kappa
e^a \omega_l^{a(k)} \nonumber \\
{\cal T}_{l+1}^{a(k)} &=& D \omega_l^{a(k)} + e_b \Sigma_l^{a(k)b} 
\end{eqnarray}
Using these curvatures the Lagrangian can be written as follows:
\begin{equation}
{\cal L} = \hat{E}_{a(2l+2)} [ c_1 {\cal R}_{l+1}^{a(l+1)b(k-l)}
{\cal R}_{l+1}^{a(l+1)b(k-l)} + c_2 {\cal T}_{l+1}^{a(l+1)b(k-l-1)}
{\cal T}_{l+1}^{a(l+1)b(k-l-1)} ]
\end{equation}
where
$$
[ c_2 - (d-k-l-1) \kappa c_1 ] = \frac{(-1)^{k-l}}{2(l+1)}
$$
Once again we face the ambiguity in the choice of the parameters. This
time it is related with the identity:
$$
\hat{E}_{a(2l+2)} [ {\cal R}^{a(l+1)}{}_{b(k-l)}
{\cal R}^{a(l+1)b(k-l)} + (d-k-l-1) \kappa 
{\cal T}^{a(l+1)}{}_{b(k-l-1)} {\cal T}^{a(l+1)b(k-l-1)} ] = 0
$$
Note that up to some difference in notation and conventions all the
results in this subsection are in agreement with the results in
\cite{Alk03}.

\section{Partially massless case in de Sitter space}

In this section we will show that combination of two fields $Y(k,l)$
and $Y(k,l-1)$ can be deformed into de Sitter space with $\kappa > 0$.
Moreover, such deformation will keep all four (appropriately modified)
gauge symmetries of these two fields.

\subsection{Lagrangian and gauge transformations}

Thus we introduce two pairs of physical and auxiliary fields: 
($\Sigma_l{}^{k+1}$, $R_k{}^l$) and ($\Omega_{l-1}^{k+1}$,
$\Phi_k{}^{l-1}$). We begin our construction with the sum of their
kinetic terms:
\begin{eqnarray}
{\cal L}_0 &=& {\cal L}_0 (\Sigma_l^{k+1}, R_k^l) + {\cal L}_0
(\Omega_{l-1}^{k+1}, \Phi_k^{l-1}) \label{lag1}
\end{eqnarray}
as well as their initial gauge transformations:
\begin{eqnarray}
\delta_0 \Sigma_l^{a(k+1)} &=& D \eta_{l-1}^{a(k+1)}, \qquad
\delta_0 R_k^{a(l)} = D \xi_{k-1}^{a(l)} + E_{b(k-l+1)}
\eta_{l-1}^{a(l)b(k-l+1)} \nonumber \\
\delta_0 \Omega_{l-1}^{a(k+1)} &=& D \eta_{l-2}^{a(k+1)}, \qquad
\delta_0 \Phi_k^{a(l-1)} = D \xi_{k-1}^{a(l-1)} + E_{b(k-l+2)}
\eta_{l-2}^{a(l-1)b(k-l+2)} \label{trans1}
\end{eqnarray}
where all derivatives are replaced by the the $dS$ covariant ones.

Now we add all possible terms with one derivative:
\begin{equation}
{\cal L}_1 = a_1 \hat{E}_{a(k+l)} \Sigma_l^{a(k+1)} \Phi_k^{a(l-1)} +
a_2 \hat{E}_{a(k+l-1)} \Omega_{l-1}^{a(k)}{}_b R_k^{a(l-1)b}
\label{lag2}
\end{equation}
and corresponding corrections to the gauge transformations:
\begin{eqnarray}
\delta_1 \Sigma_l^{a(k+1)} &=& \alpha_1 e^a e_b \eta_{l-2}^{a(k)b},
\qquad \delta_1 R_k^{a(l)} = \alpha_2 e^a \xi_{k-1}^{a(l-1)} \nonumber
\\ 
\delta_1 \Omega_{l-1}^{a(k+1)} &=& \alpha_3 \eta_{l-1}^{a(k+1)},
\qquad \quad \delta_1 \Phi_k^{a(l-1)} = \alpha_4 e_b 
\xi_{k-1}^{a(l-1)b} \label{trans2}
\end{eqnarray}
Then all variations with one derivative $\delta_0 {\cal L}_1 +
\delta_1 {\cal L}_0$ (see Appendix B for details)
cancel provided we set
\begin{equation}
\alpha_1 = \alpha_2 = - \frac{(-1)^{k-l} a_1}{l(d-k-l)}, \quad
\alpha_3 = (-1)^l a_1, \quad 
\alpha_4 = (-1)^k a_1, \quad
a_2 = (-1)^l (k+1) a_1
\end{equation}
As we have already noted, there is no possible terms quadratic in the
physical fields $R$ and $\Phi$ that can be added to the Lagrangian.
Nonetheless all variations without derivatives $\delta_1 {\cal L}_1$
indeed cancel if we put
\begin{equation}
a_1{}^2 = l (d-k-l) \kappa
\end{equation}

\subsection{Gauge invariant curvatures}

In this case we also obtain four gauge invariant curvatures:
\begin{eqnarray}
{\cal R}_{l+1}^{a(k+1)} &=& D \Sigma_l^{a(k+1)} - \alpha_1 e^a e_b
\Omega_{l-1}^{a(k)b} \nonumber \\
{\cal T}_{k+1}^{a(l)} &=& D R_k^{a(l)} + (-1)^{k-l} E_{b(k-l+1)}
\Sigma_l^{a(l)b(k-l+1)} + \alpha_2 e^a \Phi_k^{a(l-1)} \nonumber \\
{\cal R}_l^{a(k+1)} &=& D \Omega_{l-1}^{a(k+1)} - \alpha_3 
\Sigma_l^{a(k+1)} \\
{\cal T}_{k+1}^{a(l-1)} &=& D \Phi_k^{a(l-1)} - (-1)^{k-l}
E_{b(k-l+2)} \Omega_{l-1}^{a(l-1)b(k-l+2)} + \alpha_4 e_b
R_k^{a(l-1)b} \nonumber
\end{eqnarray}
By straightforward calculations one can check that these curvatures
satisfy the following differential identities:
\begin{eqnarray*}
D {\cal R}_{l+1}^{a(k+1)} &=& - \alpha_1 e^a e_b {\cal R}_l^{a(k)b} \\
D {\cal T}_{k+1}^{a(l)} &=& - E_{b(k-l+1)} 
{\cal R}_{l+1}^{a(l)b(k-l+1)} - \alpha_2 e^a {\cal T}_{l+1}^{a(l-1)} 
\\
D {\cal R}_l^{a(k+1)} &=& - \alpha_3 {\cal R}_{l+1}^{a(k+1)} \\
D {\cal T}_{l+1}^{a(l-1)} &=& - E_{b(k-l+2)} 
{\cal R}_l^{a(l-1)b(k-l+2)} - \alpha_4 e_b {\cal T}_{k+1}^{a(l-1)b}
\end{eqnarray*}
Using these curvatures one can rewrite the free Lagrangian in the
explicitly gauge invariant form:
\begin{eqnarray}
{\cal L} &=& b_1 \hat{E}_{a(2l+2)} {\cal R}_{l+1}^{a(l+1)}{}_{b(k-l)}
{\cal R}_{l+1}^{a(l+1)b(k-l)} + b_2 \hat{E}_{a(2l)}
{\cal R}_l^{a(l)}{}_{b(k-l+1)} {\cal R}_l^{a(l)b(k-l+1)} \nonumber \\
 && + b_3 \hat{E}_{a(k+l+1)} {\cal R}_l^{a(k+1)} {\cal T}_{k+1}^{a(l)}
\end{eqnarray}
where
$$
[ \frac{(l+1)^2(d-k-l-1)}{l(d-k-l)} b_1 - b_2 ] =  
\frac{a_0(k,l)}{a_1{}^2}, \qquad b_3 = - \frac{(-1)^l}{a_1}
$$
One can see that there is an ambiguity in the choice of the
coefficients $b_{1,2,3}$. This ambiguity is related with the fact that
the three terms in the Lagrangian are not independent. Indeed, using
the differential identities given above, one can show that:
\begin{eqnarray*}
0 &=& \hat{E}_{a(2l+2)} D [ {\cal R}_{l+1}^{a(l+1)}{}_{b(k-l)} 
{\cal R}_l{}^{a(l+1)b(k-l)} ] \\
 &=& - a_1 [ \frac{(l+1)^2(d-k-l-1)}{l(d-k-l)} \hat{E}_{a(2l)} 
{\cal R}_l^{a(l)}{}_{b(k-l+1)} {\cal R}_l^{a(l)b(k-l+1)} \\
 && \qquad + \hat{E}_{a(2l+2)} {\cal R}_{l+1}^{a(l+1)}{}_{b(k-l)} 
{\cal R}_{l+1}^{a(l+1)b(k-l)} ]
\end{eqnarray*}

\section{General massive case}

In this section we consider general massive case for $Y(k,l)$. By
analogy with the simple examples considered previously \cite{Zin02a},
we will use four massless fields $Y(k,l)$, $Y(k,l-1)$, $Y(k-1,l)$ and 
$Y(k-1,l-1)$.

\subsection{Lagrangian and gauge transformations}

Thus we introduce four pairs of physical and auxiliary fields: 
$(R_k{}^l, \Sigma_l{}^{k+1})$, $(h_{k-1}{}^l, \omega_l{}^k)$,
$(\Phi_k{}^{l-1}, \Omega_{l-1}{}^{k+1})$,
$(B_{k-1}{}^{l-1}, C_{l-1}{}^k)$. We begin with the sum of their
kinetic terms:
\begin{equation}
{\cal L}_0 = {\cal L}_0 (\Sigma_l^{k+1}, R_l^k) + {\cal L}_0
(\Omega_{l-1}^{k+1}, \Phi_k^{l-1}) + {\cal L}_0 
(\omega_l^k, h_{k-1}^l) + {\cal L}_0 (C_{l-1}^k, B_{k-1}^{l-1})
\end{equation}
as well as their initial gauge transformations:
\begin{eqnarray}
\delta_0 \Sigma_l^{a(k+1)} &=& D \eta_{l-1}^{a(k+1)}, \qquad
\delta_0 R_k^{a(l)} = D \xi_{k-1}^{a(l)} + E_{b(k-l+1)}
\eta_{l-1}^{a(l)b(k-l+1)} \nonumber \\
\delta_0 \Omega_{l-1}^{a(k+1)} &=& D \eta_{l-2}^{a(k+1)}, \qquad
\delta_0 \Phi_k^{a(l-1)} = D \xi_{k-1}^{a(l-1)} + E_{b(k-l+2)}
\eta_{l-2}^{a(l-1)b(k-l+2)} \nonumber \\
\delta_0 \omega_l^{a(k)} &=& D \eta_{l-1}^{a(k)}, \qquad \quad
\delta_0 h_{k-1}^{a(l)} = D \xi_{k-2}^{a(l)} + E_{b(k-l)}
\eta_{l-1}^{a(l)b(k-l)} \\
\delta_0 C_{l-1}^{a(k)} &=& D \eta_{l-2}^{a(k)}, \qquad \quad
\delta_0 B_{k-1}^{a(l-1)} = D \xi_{k-2}^{a(l-1)} + E_{b(k-l+1)} 
\eta_{l-2}^{a(l-1)b(k-l+1)} \nonumber
\end{eqnarray}
where all derivatives are $(A)dS$ covariant now.

Now we add all possible terms with one derivative (note that the
choice of the coefficients is compatible with the results in the two
previous sections):
\begin{eqnarray}
{\cal L}_1 &=& a_1 \hat{E}_{a(k+l)} \Sigma_l^{a(k+1)} \Phi_k^{a(l-1)}
+ a_2 \hat{E}_{a(k+l-1)} \Omega_{l-1}^{a(k)b} R_k^{a(l-1)b} \nonumber
\\
 && + a_3 \hat{E}_{a(k+l-1)} \Sigma_l^{a(k)b} h_{k-1}^{a(l-1)b} + a_4 
\hat{E}_{a(k+l)} \omega_l^{a(k)} R_k^{a(l)} \nonumber \\
 && + a_5 \hat{E}_{a(k+l-2)} \Omega_{l-1}^{a(k)b} B_{k-1}^{a(l-2)b} +
a_6 \hat{E}_{a(k+l-1)} C_{l-1}^{a(k)} \Phi_k^{a(l-1)} \nonumber \\
 && + a_7 \hat{E}_{a(k+l-1)} \omega_l^{a(k)} B_{k-1}^{a(l-1)} + a_8
\hat{E}_{a(k+l-2)} C_{l-1}^{a(k-1)b} h_{k-1}^{a(l-1)b} 
\end{eqnarray}
and corresponding corrections to the gauge transformations:
\begin{eqnarray}
\delta_1 \Sigma_l^{a(k+1)} &=& \alpha_1 e^a e_b \eta_{l-2}^{a(k)b} +
\alpha_5 e^a \eta_{l-1}^{a(k)} \nonumber \\
\delta_1 R_k^{a(l)} &=& \alpha_2 e^a \xi_{k-1}^{a(l-1)} + \alpha_6 e^a
e_b \xi_{k-2}^{a(l-1)b} \nonumber \\
\delta_1 \Omega_{l-1}^{a(k+1)} &=& \alpha_3 \eta_{l-1}^{a(k+1)} +
\alpha_9 e^a \eta_{l-2}^{a(k)} \nonumber \\
\delta_1 \Phi_k^{a(l-1)} &=& \alpha_4 e_b \xi_{k-1}^{a(l-1)b} +
\alpha_{10} e^a e_b \xi_{k-2}^{a(l-2)b} \nonumber \\
\delta_1 \omega_l^{a(k)} &=& \alpha_7 e_b \eta_{l-1}^{a(k)b} +
\alpha_{13} e^a e_b \eta_{l-2}^{a(k-1)b} \\
\delta_1 h_{k-1}^{a(l)} &=& \alpha_8 \xi_{k-1}^{a(l)} + \alpha_{14}
e^a \xi_{k-2}^{a(l-1)} \nonumber \\
\delta_1 C_{l-1}^{a(k)} &=& \alpha_{11} e_b \eta_{l-2}^{a(k)b} +
\alpha_{15} \eta_{l-1}^{a(k)} \nonumber \\
\delta_1 B_{k-1}^{a(l-1)} &=& \alpha_{12} \xi_{k-1}^{a(l-1)} +
\alpha_{16} e_b \xi_{k-2}^{a(l-1)b} \nonumber
\end{eqnarray}
Requirement that all the variations with one derivative $\delta_0
{\cal L}_1 + \delta_1 {\cal L}_0 = 0$ cancel gives the
same expressions for $\alpha_1 \dots \alpha_8$ as before and
$$
\alpha_9 = - \alpha_{10} = - \frac{(-1)^l a_6}{(k+1)(d-k-l+1)}, \quad
\alpha_{11} = (-1)^{k-l} a_6, \quad
\alpha_{12} = a_6
$$
$$
\alpha_{13} = \alpha_{14} = \frac{(-1)^{k-l} a_7}{l(d-k-l+1)}, \quad
\alpha_{15} = (-1)^l a_7, \quad
\alpha_{16} = - (-1)^k a_7
$$
\begin{equation}
a_2 = (-1)^l (k+1)a_1, \quad
a_3 = (-1)^l l a_4, \quad
a_5 = - (-1)^l (l-1)a_6, \quad
a_8 = (-1)^l k a_7
\end{equation}
In this case also there are no any possible terms quadratic in
physical fields that can be added to the Lagrangian. Nonetheless, all
the remaining variations vanish provided we put:
\begin{equation}
a_6{}^2 = \frac{(d-k-l+1)}{(d-k-l)} a_4{}^2, \qquad
a_7{}^2 = \frac{(d-k-l+1)}{(d-k-l)} a_1{}^2
\end{equation}
\begin{equation}
(k+1)a_1{}^2 - l a_4{}^2 = l(k+1)(d-k-l) \kappa \label{rel1}
\end{equation}
Thus all the coefficients for the terms in ${\cal L}_1$ gluing our
four massless fields together are determined by the two main ones
$a_1$ and $a_4$ (as schematically shown on Fig.1) that satisfy the
relation (\ref{rel1}).
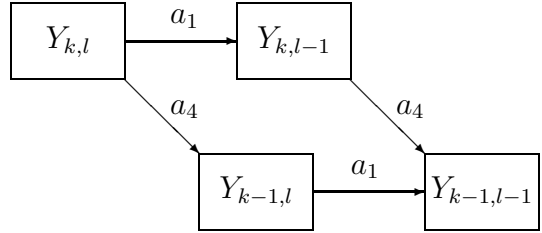
\begin{figure}[htb]
\begin{center}
\begin{picture}(90,32)
\put(10,21){\framebox(15,10)[]{$Y_{k,l}$}}
\put(40,21){\framebox(15,10)[]{$Y_{k,l-1}$}}
\put(35,1){\framebox(15,10)[]{$Y_{k-1,l}$}}
\put(65,1){\framebox(15,10)[]{$Y_{k-1,l-1}$}}
\put(25,26){\vector(1,0){15}}
\put(30,26){\makebox(6,6)[]{$a_1$}}
\put(25,21){\vector(1,-1){10}}
\put(30,14){\makebox(6,6)[]{$a_4$}}
\put(55,21){\vector(1,-1){10}}
\put(60,14){\makebox(6,6)[]{$a_4$}}
\put(50,6){\vector(1,0){15}}
\put(54,6){\makebox(6,6)[]{$a_1$}}
\end{picture}
\end{center}
\caption{General massive theory for $Y(k,l)$ field}
\end{figure}
From the relation (\ref{rel1}) it follows that in the de Sitter space
($\kappa > 0$) we can take a limit $a_4 \to 0$. In this case the whole
system decomposes in two independent subsystems (see Fig.2) both of
them corresponding to partially massless case considered in Section 3.
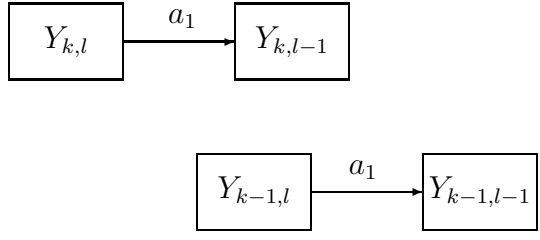
\begin{figure}[htb]
\begin{center}
\begin{picture}(90,32)
\put(10,21){\framebox(15,10)[]{$Y_{k,l}$}}
\put(40,21){\framebox(15,10)[]{$Y_{k,l-1}$}}
\put(35,1){\framebox(15,10)[]{$Y_{k-1,l}$}}
\put(65,1){\framebox(15,10)[]{$Y_{k-1,l-1}$}}
\put(25,26){\vector(1,0){15}}
\put(30,26){\makebox(6,6)[]{$a_1$}}
\put(50,6){\vector(1,0){15}}
\put(54,6){\makebox(6,6)[]{$a_1$}}
\end{picture}
\end{center}
\caption{Partially massless limit in de Sitter space}
\end{figure}

From the other side, in the anti de Sitter space ($\kappa < 0$) we can
take a limit $a_1 \to 0$ where our system also decomposes into two
independent subsystems (see Fig.3), each of them corresponding to the
partially massless case considered in Section 2.
\begin{figure}[htb]
\begin{center}
\begin{picture}(90,32)
\put(10,21){\framebox(15,10)[]{$Y_{k,l}$}}
\put(40,21){\framebox(15,10)[]{$Y_{k,l-1}$}}
\put(35,1){\framebox(15,10)[]{$Y_{k-1,l}$}}
\put(65,1){\framebox(15,10)[]{$Y_{k-1,l-1}$}}
\put(25,21){\vector(1,-1){10}}
\put(30,14){\makebox(6,6)[]{$a_4$}}
\put(55,21){\vector(1,-1){10}}
\put(60,14){\makebox(6,6)[]{$a_4$}}
\end{picture}
\end{center}
\caption{Partially massless limit in anti de Sitter space}
\end{figure}
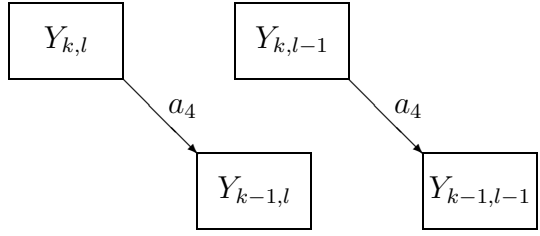

\subsection{Partial gauge fixing}

In this case there exist as many as eight gauge invariant curvatures:
\begin{eqnarray}
{\cal R}_{l+1}^{a(k+1)} &=& D \Sigma_l^{a(k+1)} - \alpha_1 e^a e_b 
\Omega_{l-1}^{a(k)b} + \alpha_5 e^a \omega_l^{a(k)} \nonumber  \\
{\cal T}_{k+1}^{a(l)} &=& D R_k^{a(l)} + (-1)^{k-l} E_{b(k-l+1)}
\Sigma_l^{a(l)b(k-l+1)} + \alpha_2 e^a \Phi_k^{a(l-1)} - \alpha_6 e^a
e_b h_{k-1}^{a(l-1)b} \nonumber \\
{\cal R}_l^{a(k+1)} &=& D \Omega_{l-1}^{a(k+1)} - \alpha_3 
\Sigma_l^{a(k+1)} + \alpha_9 e^a C_{l-1}^{a(k)} \nonumber  \\
{\cal T}_{k+1}^{a(l-1)} &=& D \Phi_k^{a(l-1)} - (-1)^{k-l} 
E_{b(k-l+2)} \Omega_{l-1}^{a(l-1)b(k-l+2)} + \alpha_4 e_b
R_k^{a(l-1)b} - \alpha_{10} e^a e_b B_{k-1}^{a(l-2)b} \nonumber \\
{\cal R}_{l+1}^{a(k)} &=& D \omega_l^{a(k)} + \alpha_7 e_b
\Sigma_l^{a(k)b} - \alpha_{13} e^a e_b C_{l-1}^{a(k-1)b}  \\
{\cal T}_k^{a(l)} &=& D h_{k-1}^{a(l)} - (-1)^{k-l} E_{b(k-l)} 
\omega_l^{a(l)b(k-l)} - \alpha_8 R_k^{a(l)} + \alpha_{14} e^a 
B_{k-1}^{a(l-1)} \nonumber \\
{\cal R}_l^{a(k)} &=& D C_{l-1}^{a(k)} + \alpha_{11} e_b
\Omega_{l-1}^{a(k)b} - \alpha_{15} \omega_l^{a(k)} \nonumber \\
{\cal T}_k^{a(l-1)} &=& D B_{k-1}^{a(l-1)} + (-1)^{k-l} E_{b(k-l+1)}
C_{l-1}^{a(l-1)b(k-l+1)} - \alpha_{12} \Phi_k^{a(l-1)} + \alpha_{16}
e_b h_{k-1}^{a(l-1)b} \nonumber
\end{eqnarray}
In principle, it is possible to rewrite the free Lagrangian in terms
of these curvatures, but here one can construct nine invariants
quadratic in curvatures as well as four identities that these
invariants satisfy. So as in the Section 2 we proceed with partial
gauge fixing. Namely, we use $\xi_{k-1}^{a(l)}$ and 
$\xi_{k-1}^{a(l-1)}$ transformations to set $h_{k-1}^{a(l)} = 0$ and 
$B_{k-1}^{a(l-1)} = 0$. Then we solve the on-shell constraints 
${\cal T}_l^{a(l)} \approx 0$ and ${\cal T}_k^{a(l-1)} \approx 0$ and
get:
\begin{eqnarray*}
R_k^{a(l)} &=& \frac{(-1)^{k-l}}{a_4} E_{b(k-l)} \omega_l^{a(l)b(k-l)}
\\
\Phi_k^{a(l-1)} &=& \frac{(-1)^{k-l}}{a_6} E_{b(k-l+1)} 
C_{l-1}^{a(l-1)b(k-l+1)} 
\end{eqnarray*}
Taking into account that the fields $\omega$ and $C$ will now play the
roles of physical (and not the auxiliary) ones, we make appropriate
re-scaling:
$$
\omega_l^{a(k)} \Rightarrow (-1)^{k-l} a_4 \omega_l^{a(k)}, \qquad
C_{l-1}^{a(k)} \Rightarrow (-1)^{k-l} a_6 C_{l-1}^{a(k)}
$$
These leaves with the set of four fields with the Lagrangian:
\begin{eqnarray*}
{\cal L} &=& \frac{(-1)^{k-l}(l+1)}{2} \hat{E}_{a(2l)}
\Sigma_l^{a(l)}{}_{b(k-l+1)} \Sigma_l^{a(l)b(k-l+1)} +
\hat{E}_{a(2l+1)} \Sigma_l^{a(l+1)}{}_{b(k-l)} D
\omega_l^{a(l)b(k-l)} \\
 && + \frac{(-1)^{k-l}l(l+1)}{2} \hat{E}_{a(2l-2)}
\Omega_{l-1}^{a(l-1)}{}_{k-l+2)} \Omega_{l-1}^{a(l-1)b(k-l+2)} \\
 && \qquad + (l+1) \hat{E}_{a(2l-1)} \Omega_{l-1}^{a(l)}{}_{b(k-l+1)}
D C_{l-1}^{a(l-1)b(k-l+1)} \\
 && + (l+1) a_1 \hat{E}_{a(2l-1)} [ \Sigma_l^{a(l)}{}_{b(k-l+1)}
C_{l-1}^{a(l-1)b(k-l+1)} + (-1)^k \Omega_{l-1}^{a(l)}{}_{b(k-l+1)}
\omega_l^{a(l_1)b(k-l+1)} ] \\
 && + \frac{(-1)^{k-l}(l+1)}{2(k+1)} [ a_4{}^2 \hat{E}_{a(2l)}
\omega_l^{a(l)}{}_{b(k-l)} \omega_l^{a(l)b(k-l)} + l a_6{}^2
\hat{E}_{a(2l-2)} C_{l-1}^{a(l-1)}{}_{b(k-l+1)} 
C_{l-1}^{a(l-1)b(k-l+1)} ]
\end{eqnarray*}
which is still invariant under the remaining gauge transformations:
\begin{eqnarray}
\delta \Sigma_l^{a(k+1)} &=& D \eta_{l-1}^{a(k+1)} + \alpha_1 e^a e_b 
\eta_{l-2}^{a(k)b} + \tilde{\alpha}_5 e^a \eta_{l-1}^{a(k)} \nonumber
\\
\delta \omega_l^{a(k)} &=& D \eta_{l-1}^{a(k)} + e_b
\eta_{l-1}^{a(k)b} - \alpha_1 e^a e_b
\eta_{l-2}^{a(k-1)b} \nonumber \\
\delta \Omega_{l-1}^{a(k+1)} &=& D \eta_{l-2}^{a(k+1)} + \alpha_3 
\eta_{l-1}^{a(k+1)} + \tilde{\alpha}_5 e^a \eta_{l-2}^{a(k)} \\
\delta C_{l-1}^{a(k)} &=& D \eta_{l-2}^{a(k)} + e_b 
\eta_{l-2}^{a(k)b} + \alpha_3 \eta_{l-1}^{a(k)} \nonumber
\end{eqnarray}
where
$$
\tilde{\alpha}_5 = - \frac{(-1)^k a_4{}^2}{(k+1)(d-k-l)}
$$
and the new set of curvatures
\begin{eqnarray}
{\cal R}_{l+1}^{a(k+1)} &=& D \Sigma_l^{a(k+1)} - \alpha_1 e^a e_b 
\Omega_{l-1}^{a(k)b} + \tilde{\alpha}_5 e^a \omega_l^{a(k)} \nonumber
\\
{\cal T}_{l+1}^{a(k)} &=& D \omega_l^{a(k)} + e_b
\Sigma_l^{a(k)b} + \alpha_1 e^a e_b C_{l-1}^{a(k-1)b} \nonumber \\
{\cal R}_l^{a(k+1)} &=& D \Omega_{l-1}^{a(k+1)} - \alpha_3 
\Sigma_l^{a(k+1)} + \tilde{\alpha}_5 e^a C_{l-1}^{a(k)}  \\
{\cal T}_l^{a(k)} &=& D C_{l-1}^{a(k)} + e_b \Omega_{l-1}^{a(k)b} -
\alpha_3 \omega_l^{a(k)} \nonumber
\end{eqnarray}
Trying to rewrite this Lagrangian it terms of curvatures, we found
that there are six invariants quadratic in them as well as four
identities they satisfy. The simplest example we managed to find looks
like:
\begin{equation}
{\cal L} = c_1 \hat{E}_{a(2l+2)} {\cal T}_{l+1}^{a(l+1)b(k-l-1)}
{\cal T}_{l+1}^{a(l+1)b(k-l-1)} + c_2 \hat{E}_{a(2l)}
{\cal T}_l^{a(l)b(k-l)} {\cal T}_l^{a(l)b(k-l)}
\end{equation}
where
$$
c_1 = \frac{(-1)^{k-l}}{2(l+1)}, \qquad
c_2 = \frac{(-1)^{k-l}(l+1)}{2l}
$$

\section{Special case $Y(k,k)$}

In this section we consider massive version for the special case
$Y(k,k)$. Similarly to the example $Y(2,2)$ considered previously
\cite{Zin02a} we will need three massless fields $Y(k,k)$, $Y(k,k-1)$
and $Y(k-1,k-1)$ only.

\subsection{Lagrangian and gauge transformations}

We introduce three pairs of physical and auxiliary fields:
$(R_k{}^k,\Sigma_k{}^{k+1})$, 
$(\Phi_k{}^{k-1},\Omega_{k-1}{}^{k+1})$, 
$(h_{k-1}{}^{k-1},\omega_{k-1}{}^k)$ and begin with the sum of their
kinetic terms:
\begin{equation}
{\cal L}_0 = {\cal L}_0 (\Sigma_k^{k+1}, R_k^k) + {\cal L}_0
(\Omega_{k-1}^{k+1}, \Phi_k^{k-1}) + {\cal L}_0 
(\omega_{k-1}^k, h_{k-1}^{k-1}) 
\end{equation}
as well as their initial gauge transformations:
\begin{eqnarray}
\delta_0 \Sigma_k^{a(k+1)} &=& D \eta_{k-1}^{a(k+1)}, \qquad
\delta_0 R_k^{a(k)} = D \xi_{k-1}^{a(k)} + e_b \eta_{k-1}^{a(k)b}
\nonumber \\
\delta_0 \Omega_{k-1}^{a(k+1)} &=& D \eta_{k-2}^{a(k+1)}, \qquad
\delta_0 \Phi_k^{a(k-1)} = D \xi_{k-1}^{a(k-1)} + e_b e_c
\eta_{k-2}^{a(k-1)bc} \\
\delta_0 \omega_{k-1}^{a(k)} &=& D \eta_{k-2}^{a(k)}, \qquad
\delta_0 h_{k-1}^{a(k-1)} = D \xi_{k-2}^{a(k-1)} + e_b 
\eta_{k-2}^{a(k-1)b} \nonumber
\end{eqnarray}

Now we add all possible terms with one derivative:
\begin{eqnarray}
{\cal L}_1 &=& a_1 \hat{E}_{a(2k)} \Sigma_k^{a(k+1)} \Phi_k^{a(k-1)} +
a_2 \hat{E}_{a(2k-1)} \Omega_{k-1}^{a(k)}{}_b R_k^{a(k-1)b} \nonumber
\\
 && + a_3 \hat{E}_{a(2k-2)} \Omega_{k-1}^{a(k)}{}_b h_{k-1}^{a(k-2)b}
+ a_4 \hat{E}_{a(2k-1)} \omega_{k-1}^{a(k)} \Phi_k^{a(k-1)}
\end{eqnarray}
and corresponding corrections to the gauge transformations:
\begin{eqnarray}
\delta_1 \Sigma_k^{a(k+1)} &=& \alpha_1 e^a e_b \eta_{k-2}^{a(k)b},
\qquad \delta_1 R_k^{a(k)} = \alpha_2 e^a \xi_{k-1}^{a(k-1)} \nonumber
\\
\delta_0 \Omega_{k-1}^{a(k+1)} &=& \alpha_3 \eta_{k-1}^{a(k+1)} +
\alpha_4 e^a \eta_{k-2}^{a(k)} \nonumber \\
\delta_1 \Phi_k^{a(k-1)} &=& \alpha_5 e_b \xi_{k-1}^{a(k-1)b} +
\alpha_6 e^a e_b \xi_{k-2}^{a(k-2)b} \\
\delta_1 \omega_{k-1}^{a(k)} &=& \alpha_7 e_b \eta_{k-2}^{a(k)b},
\qquad \delta_1 h_{k-1}^{a(k-1)} = \alpha_8 \xi_{k-1}^{a(k-1)}
\nonumber
\end{eqnarray}
First of all we calculate all variations with one derivative. We found
that all of them vanish provided we set
$$
\alpha_1 = \alpha_2 = \frac{a_1}{k(d-2k)}, \qquad
\alpha_3 = \alpha_5 = - (-1)^k a_1
$$
$$
\alpha_4 = - \alpha_6 = \frac{(-1)^k a_4}{(k+1)(d-2k+1)}, \qquad
\alpha_7 = \alpha_8 = - a_4
$$
\begin{equation}
a_2 = (-1)^k (k+1) a_1, \qquad
a_3 = - (-1)^k (k-1) a_4
\end{equation}
But in this case it is not possible to cancel the remaining variations
just by adjusting the two main parameters $a_1$ and $a_4$. Happily, we
have two physical fields $R_k^{a(k)}$ and $h_{k-1}^{a(k-1)}$ with
equal numbers of world and local indices, so we proceed and introduce
terms without derivatives:
\begin{equation}
{\cal L}_2 = b_1 \hat{E}_{a(2k)} R_k^{a(k)} R_k^{a(k)} + b_2
\hat{E}_{a(2k-1)} R_k^{a(k)} h_{k-1}^{a(k-1)} + b_3 \hat{E}_{a(2k-2)}
h_{k-1}^{a(k-1)} h_{k-1}^{a(k-1)}
\end{equation}
and corresponding corrections to the gauge transformations:
\begin{eqnarray}
\delta_2 \Sigma_k^{a(k+1)} &=& \beta_1 e^a \xi_{k-1}^{a(k)} + \beta_2
e^a e^a \xi_{k-2}^{a(k-1)} \nonumber \\
\delta_2 \omega_{k-1}^{a(k)} &=& \beta_3 \xi_{k-1}^{a(k)} + \beta_4
e^a \xi_{k-2}^{a(k-1)}
\end{eqnarray}
For this corrections to be consistent with ${\cal L}_2$ we have to
put:
$$
\beta_1 = \frac{(-1)^k 2b_1}{(k+1)(d-2k)}, \qquad
\beta_2 = - \frac{(-1)^k b_2}{k(k+1)(d-2k)(d-2k+1)}
$$
$$
\beta_3 = - (-1)^k b_2, \qquad \qquad
\beta_4 = \frac{(-1)^k 2b_3}{k(d-2k+2)}
$$
Now we require that all variations vanish and obtain:
\begin{eqnarray}
2kb_1 &=& - (k+1) a_1{}^2 + k(k+1)(d-2k) \kappa, \qquad
b_2 = 	- a_1a_4 \nonumber \\
2b_3 &=& - \frac{k(d-2k+2)}{(k+1)(d-2k+1)} a_4{}^2 - k(d-2k+2)\kappa
\end{eqnarray}
\begin{equation}
(k+1)(d-2k+1) a_1{}^2 - k(d-2k) a_4{}^2 = k(k+1)(d-2k)(d-2k+1) \kappa
\label{rel2}
\end{equation}

Thus all the coefficients that determine the mixing our three massless
fields together are determined by the two main ones $a_1$ and $a_4$
(see Fig.4) that in turn satisfy the relation (\ref{rel2}).
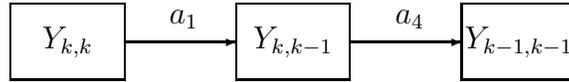
\begin{figure}[htb]
\begin{center}
\begin{picture}(95,12)
\put(10,1){\framebox(15,10)[]{$Y_{k,k}$}}
\put(40,1){\framebox(15,10)[]{$Y_{k,k-1}$}}
\put(70,1){\framebox(15,10)[]{$Y_{k-1,k-1}$}}
\put(25,6){\vector(1,0){15}}
\put(30,6){\makebox(6,6)[]{$a_1$}}
\put(55,6){\vector(1,0){15}}
\put(60,6){\makebox(6,6)[]{$a_4$}}
\end{picture}
\end{center}
\caption{General massive case for the $Y(k,k)$ field}
\end{figure}

The relation (\ref{rel2}) shows that in the de Sitter space 
($\kappa > 0$) we can take a limit $a_4 \to 0$. In this limit our
system decomposes into two independent subsystems (see Fig.5) where the
first one corresponds to the partially massless case considered in
Section 3.
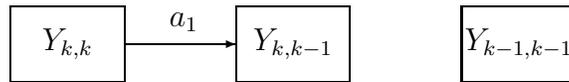
\begin{figure}[htb]
\begin{center}
\begin{picture}(95,12)
\put(10,1){\framebox(15,10)[]{$Y_{k,k}$}}
\put(40,1){\framebox(15,10)[]{$Y_{k,k-1}$}}
\put(70,1){\framebox(15,10)[]{$Y_{k-1,k-1}$}}
\put(25,6){\vector(1,0){15}}
\put(30,6){\makebox(6,6)[]{$a_1$}}
\end{picture}
\end{center}
\caption{Partially massless limit in de Sitter space}
\end{figure}

From the other hand, in the anti de Sitter space ($\kappa < 0$) we can
take a limit $a_1 \to 0$. Then the whole system also decomposes into
two independent subsystems (see Fig.6) where the first one is just the
massless $Y(k,k)$ field as in Section 1, while the second one
corresponds to partially massless case considered in Section 2.
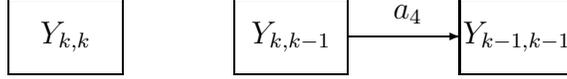
\begin{figure}[htb]
\begin{center}
\begin{picture}(95,12)
\put(10,1){\framebox(15,10)[]{$Y_{k,k}$}}
\put(40,1){\framebox(15,10)[]{$Y_{k,k-1}$}}
\put(70,1){\framebox(15,10)[]{$Y_{k-1,k-1}$}}
\put(55,6){\vector(1,0){15}}
\put(60,6){\makebox(6,6)[]{$a_4$}}
\end{picture}
\end{center}
\caption{Massless limit in anti de Sitter space}
\end{figure}

\subsection{Curvatures and partial gauge fixing}

This time we have six gauge invariant curvatures:
\begin{eqnarray*}
{\cal R}_{k+1}^{a(k+1)} &=& D \Sigma_k^{a(k+1)} - \alpha_1 e^a e_b
\Omega_{k-1}^{a(k)b} + \beta_1 e^a R_k^{a(k)} - \beta_2 e^{a(2)}
h_{k-1}^{a(k-1)} \\
{\cal T}_{k+1}^{a(k)} &=& D R_k^{a(k)} + e_b \Sigma_k^{a(k)b} +
\alpha_2 e^a \Phi_k^{a(k-1)}  \\
{\cal R}_k^{a(k+1)} &=& D \Omega_{k-1}^{a(k+1)} - \alpha_3 
\Sigma_k^{a(k+1)} + \alpha_4 e^a \omega_{k-1}^{a(k)}    \\
{\cal T}_{k+1}^{a(k-1)} &=& D \Phi_k^{a(k-1)} - E_{b(2)}
\Omega_{k-1}^{a(k-1)b(2)} + \alpha_5 e_b R_k^{a(k-1)b} - \alpha_6 e^a
e_b h_{k-1}^{a(k-1)b}  \\
{\cal R}_k^{a(k)} &=& D \omega_{k-1}^{a(k)} + \alpha_7 e_b
\Omega_{k-1}^{a(k)b} - \beta_3 R_k^{a(k)} + \beta_4 e^a 
h_{k-1}^{a(k-1)} \\
{\cal T}_k^{a(k-1)} &=& D h_{k-1}^{a(k-1)} + e_b 
\omega_{k-1}^{a(k-1)b} - \alpha_8 \Phi_k^{a(k-1)}
\end{eqnarray*}
Again to simplify further investigations we proceed with the partial
gauge fixing. Namely, we use $\xi_{k-1}^{a(k-1)}$ gauge
transformations to fix $h^{a(k-1)} = 0$. Then solving the on-shell
constraint ${\cal T}_k^{a(k-1)} = 0$ we obtain:
$$
\Phi^{a(k-1)} = - \frac{1}{a_4} e_b \omega^{a(k-1)b}
$$
Again we make the appropriate re-scaling
$$
\omega^{a(k)} \Rightarrow a_4 \omega^{a(k)}
$$
Then for the remaining four fields we obtain the following Lagrangian:
\begin{eqnarray}
{\cal L} &=& - \frac{(k+1)}{2} \hat{E}_{a(2k)} \Sigma_k^{a(k)}{}_b
\Sigma_k^{a(k)b} - \hat{E}_{a(2k+1)} \Sigma_k^{a(k+1)} D R_k^{a(k)}
\nonumber \\
 && - \frac{k(k+1)}{2} \hat{E}_{a(2k-2)} 
\Omega_{k-1}^{a(k-1)}{}_{b(2)} \Omega_{k-1}^{a(k-1)b(2)} - (k+1)
\hat{E}_{a(2k-1)} \Omega_{k-1}^{a(k)}{}_b D  \omega_{k-1}^{a(k-1)b}
\nonumber \\
 && + (k+1)a_1 \hat{E}_{a(2k-1)} [ - \Sigma_k^{a(k)}{}_b
\omega_{k-1}^{a(k-1)b} + (-1)^k \Omega_{k-1}^{a(k)}{}_b R_k^{a(k-1)b}
] \nonumber \\
 && +  b_1 \hat{E}_{a(2k)} R_k^{a(k)} R_k^{a(k)} 
 + \frac{k(d-2k+1)b_1}{(d-2k)} \hat{E}_{a(2k-2)} 
\omega_{k-1}^{a(k-1)}{}_b \omega_{k-1}^{a(k-1)b} 
\end{eqnarray}
which is still invariant under the following gauge transformations: 
\begin{eqnarray}
\delta \Sigma_k^{a(k+1)} &=& D \eta_{k-1}^{a(k+1)} + \alpha_1 e^a e_b
\eta_{k-2}^{a(k)b} + \beta_1 e^a \xi_{k-1}^{a(k)} \nonumber \\
\delta R_k^{a(k)} &=& D \xi_{k-1}^{a(k)} + e_b \eta_{k-1}^{a(k)b} +
\alpha_1 e^a e_b \eta_{k-2}^{a(k-1)b} \nonumber \\
\delta \Omega_{k-1}^{a(k+1)} &=& D \eta_{k-2}^{a(k+1)} - (-1)^k a_1 
\eta_{k-1}^{a(k+1)} - \beta_1 e^a \eta_{k-2}^{a(k)} \\
\delta \omega_{k-1}^{a(k)} &=& D \eta_{k-2}^{a(k)} - e_b 
\eta_{k-2}^{a(k)b} + (-1)^k a_1 \xi_{k-1}^{a(k)} \nonumber
\end{eqnarray}
Also we can construct a set of the new gauge invariant curvatures:
\begin{eqnarray}
{\cal R}_{k+1}^{a(k+1)} &=& D \Sigma_k^{a(k+1)} - \alpha_1 e^a e_b
\Omega_{k-1}^{a(k)b} + \beta_1 e^a R_k^{a(k)} \nonumber \\
{\cal T}_{k+1}^{a(k)} &=& D R_k^{a(k)} + e_b \Sigma_k^{a(k)b} 
- \alpha_1 e^a e_b \omega_{k-1}^{a(k-1)b} \nonumber \\
{\cal R}_k^{a(k+1)} &=& D \Omega_{k-1}^{a(k+1)} + (-1)^k a_1
\Sigma_k^{a(k+1)} - \beta_1 e^a \omega_{k-1}^{a(k)}    \\
{\cal T}_k^{a(k)} &=& D \omega_{k-1}^{a(k)} - e_b \Omega_{k-1}^{a(k)b}
- (-1)^k a_1 R_k^{a(k)}  \nonumber
\end{eqnarray}
Trying to rewrite this Lagrangian in terms of curvatures we have found
that there exist five invariants quadratic in them as well as two
identities these invariants satisfy. The simplest solution solution we
managed to find is:
\begin{equation}
{\cal L} = c_1 \hat{E}_{a(2k+2)} {\cal R}_{k+1}^{a(k+1)}
{\cal R}_{k+1}^{a(k+1)} + c_2 \hat{E}_{a(2k)}
{\cal R}_k^{a(k)}{}_b {\cal R}_k^{a(k)b}
\end{equation}
where
$$
c_1 = \frac{(d-2k)}{4(d-2k-1)b_1}, \qquad
c_2 = \frac{(k+1)^2}{4kb_1}
$$

\section*{Acknowledgments}

Work supported in parts by RFBR grant 14-02-01172.

\appendix

\section{Partially massless case in anti de Sitter space}

Non-invariance due to non-commutativity of $AdS$ covariant
derivatives:
\begin{eqnarray*}
\delta_0 {\cal L}_0 &=& - l(k+1)(d-k-l)\kappa \hat{E}_{a(k+l-1)}
\eta_{l-1}^{a(k)b} R_k^{a(l-1)}{}_b \\
 && - (-1)^l l(k+1)(d-k-l)\kappa \hat{E}_{a(k+l-1)} 
\Sigma_l^{a(k)}{}_b \xi_{k-1}^{a(l-1)b} \\
 && - lk(d-k-l+1)\kappa \hat{E}_{a(k+l-2)} \eta_{l-1}^{a(k-1)b}
h_{k-1}^{a(l-1)}{}_b \\
 && - (-1)^l lk(d-k-l+1)\kappa \hat{E}_{a(k+l-2)} 
\omega_l^{a(k-1)}{}_b \xi_{k-2}^{a(l-1)b}
\end{eqnarray*}
Requirement the $\delta_1 {\cal L}_0 + \delta_0 {\cal L}_1 = 0$ gives
$$
\alpha_5 = - \alpha_6 = - \frac{a_3}{l(k+1)(d-k-l)}, \quad
\alpha_7 = (-1)^{k-l} a_4, \quad
\alpha_8 = - a_4, \quad
a_3 = (-1)^l l a_4
$$
Contributions from $\delta_1 {\cal L}_1$
\begin{eqnarray*}
\delta_1 {\cal L}_1 &=& - (-1)^{k-l} la_4\alpha_7 \hat{E}_{a(k+l-1)}
\eta_{l-1}^{a(k)b} R_k^{a(l-1)}{}_b \\
 && + a_3\alpha_8 \hat{E}_{a(k+l-1)} \Sigma_l^{a(k)}{}_b
\xi_{k-1}^{a(l-1)b} \\
 && + k(d-k-l+1) a_3\alpha_5 \hat{E}_{a(k+l-2)} \eta_{l-1}^{a(k-1)b}
h_{k-1}^{a(l-1)}{}_b  \\
 && - lk (d-k-l+1) a_4\alpha_6 \hat{E}_{a(k+l-2)}
\omega_l^{a(k-1)}{}_b \xi_{k-2}^{a(l-1)b}
\end{eqnarray*}
Requirement $\delta_1 {\cal L}_1 + \delta_0 {\cal L}_0 = 0$ gives
$$
a_4{}^2 = - (k+1)(d-k-l)\kappa
$$

\section{Partially massless case in de Sitter space}

Non-invariance due to non-commutativity of $AdS$ covariant
derivatives:
\begin{eqnarray*}
\delta_0 {\cal L}_0 &=& - l(k+1)(d-k-l)\kappa \hat{E}_{a(k+l-1)}
\eta_{l-1}^{a(k)b} R_k^{a(l-1)}{}_b  \\
 && + (l-1)(k+1)(d-k-l+1)\kappa \hat{E}_{a(k+l-2)} \eta_{l-2}^{a(k)b}
\Phi_k^{a(l-2)}{}_b  \\
 && - (-1)^l l(k+1)(d-k-l)\kappa \hat{E}_{a(k+l-1)} 
\Sigma_l^{a(k)}{}_b \xi_{k-1}^{a(l-1)b} \\
 && - (-1)^l (l-1)(k+1)(d-k-l+1)\kappa \hat{E}_{a(k+l-2)}
\Omega_{l-1}^{a(k)}{}_b \xi_{k-2}^{a(l-2)b}
\end{eqnarray*}
where ${\cal L}_0$ given by (\ref{lag1}) and $\delta_0$ by
(\ref{trans1}). Requirement that $\delta_1 {\cal L}_0 + \delta_0 
{\cal L}_1 = 0$ gives
$$
\alpha_1 = \alpha_2 = - \frac{(-1)^{k-l} a_1}{l(d-k-l)}, \quad
\alpha_3 = (-1)^l a_1, \quad 
\alpha_4 = (-1)^k a_1, \quad
a_2 = (-1)^l (k+1) a_1
$$
Contributions from $\delta_1 {\cal L}_1$:
\begin{eqnarray*}
\delta_1 {\cal L}_1 &=& a_2\alpha_3 \hat{E}_{a(k+l-1)}
\eta_{k-1}^{a(k)b} R_k^{a(l-1)}{}_b  \\
 && + (-1)^{k-l} (l-1)(k+1)(d-k-l+1)a_1\alpha_1 \hat{E}_{a(k+l-2)}
\eta_{l-2}^{a(k)b} \Phi_k^{a(l-2)}{}_b \\
 && + (-1)^{k-l} (k+1)a_1\alpha_4 \hat{E}_{a(k+l-1)} 
\Sigma_l^{a(k)}{}_b \xi_{k-1}^{a(l-1)b}  \\
 && - (-1)^{k-l} (l-1)(d-k-l+1)a_2\alpha_2
\hat{E}_{a(k+l-2)} \Omega_{l-1}^{a(k)}{}_b \xi_{k-1}^{a(l-2)b}
\end{eqnarray*}
Requirement $\delta_0 {\cal L}_0 + \delta_1 {\cal L}_1 = 0$ gives
$$
a_1{}^2 = l (d-k-l) \kappa
$$

\end{document}